%% file: sigconf_camera_ready.tex
\documentclass[sigconf]{acmart}

\AtBeginDocument{%
  }

\copyrightyear{2027}
\acmYear{2027}
\setcopyright{cc}
\setcctype{by}
\acmConference[WSDM '27]{Proceedings of the Twentieth ACM International Conference on Web Search and Data Mining}{February 15--19, 2027}{Hong Kong, Hong Kong}
\acmBooktitle{Proceedings of the Twentieth ACM International Conference on Web Search and Data Mining (WSDM '27), February 15--19, 2027, Hong Kong, Hong Kong}
\acmDOI{10.1145/3862322.3863634}
\acmISBN{979-8-4007-3117-4/2027/02}

\acmSubmissionID{1884}

\usepackage{yao}
\usepackage{svg}
\usepackage{colortbl}
\definecolor{oursrow}{RGB}{232,242,249}
\definecolor{controlrow}{RGB}{246,246,246}
\begin{document}

\title{From Interests to Semantic IDs:\allowbreak{} Retrieval-Grounded\allowbreak{} Credit Assignment for Generative Recommendation}

\author{Mengdan Zhu}
\authornote{Work done during an internship at Microsoft.}
\affiliation{%
  \institution{Emory University}
  \city{Atlanta}
     \state{GA}
     \country{USA}
}
\email{mengdan.zhu@emory.edu}

\author{Yufan Zhao}
\affiliation{%
  \institution{Microsoft}
  \city{Redmond}
     \state{WA}
     \country{USA}
}
\email{yufzhao@microsoft.com}

\author{Yao Zhao}
\affiliation{%
  \institution{Microsoft}
  \city{Redmond}
     \state{WA}
     \country{USA}
}
\email{yaozhao2@microsoft.com}

\author{Sophie Di}
\affiliation{%
  \institution{Cornell University}
  \city{Ithaca}
     \state{NY}
     \country{USA}
}
\email{szd5@cornell.edu}

\author{Tao Di}
\affiliation{%
  \institution{Microsoft}
  \city{Redmond}
     \state{WA}
     \country{USA}
}
\email{taodi@microsoft.com}

\author{Yulan Yan}
\affiliation{%
  \institution{Microsoft}
  \city{Redmond}
     \state{WA}
     \country{USA}
}
\email{yulanyan@microsoft.com}

\author{Sridhar Iyer}
\affiliation{%
  \institution{Microsoft}
  \city{Redmond}
     \state{WA}
     \country{USA}
}
\email{sridhariyer@microsoft.com}

\author{Liang Zhao}
\affiliation{%
  \institution{Emory University}
  \city{Atlanta}
     \state{GA}
     \country{USA}
}
\email{liang.zhao@emory.edu}

\renewcommand{\shortauthors}{Mengdan Zhu et al.}

\begin{abstract}

Semantic IDs (SIDs) encode each catalog item as a short token sequence, enabling generative recommenders to predict the next item autoregressively. Reasoning-enhanced variants, an increasingly common extension, first generate a textual trace and then decode a next-item SID by beam search. Such recommenders are commonly trained with group-relative policy optimization under an exact-match SID reward, which is sparse in large catalogs. Two failure modes follow. When all rollouts in a group miss the target, the group yields zero advantage and no learning signal. Rollouts sharing the same SID reward receive identical advantages, however much their traces differ. In both cases the reward reflects only the decoded SID, never the reasoning that produced it. This creates a credit-assignment gap.

We address this gap with retrieval-grounded query attribution\footnote{Code available at
\url{https://github.com/YuFan-Microsoft/Retrieval-Grounded-Credit-Assignment-for-Generative-Recommendation}.}. Each trace is structured into a history summary, a set of interest hypotheses, and a final SID. A frozen retriever executes every hypothesis as a catalog query, so that each hypothesis becomes independently verifiable rather than judged only through the final SID. A rollout is rewarded when any of its queries retrieves the target within the \mbox{top-$K$}, and per-query hit indicators localize that reward to individual hypotheses. Credit is thus assigned at the span level: only hypotheses that individually hit receive positive retrieval advantage, while the retrieval channel never updates the final SID span. Rollouts that share a SID reward can therefore receive different updates. Across experiments on three Amazon Reviews datasets, this yields consistent improvements in SID recommendation.
On Video Games, an oracle analysis further reveals the potential of interest-conditioned SID decoding: selecting the target-relevant query among generated interests improves both recall and ranking.
\end{abstract}



\setlength{\emergencystretch}{1em}
\maketitle
\setlength{\emergencystretch}{0pt}

\input{sections/introduction}
\input{sections/related_work}
\input{sections/methodology}
\input{sections/experiments_arxiv}

\input{sections/conclusion}
\input{sections/ethical_considerations}
\bibliographystyle{ACM-Reference-Format}
\bibliography{references}

\end{document}

%% file: sections/introduction.tex
\section{Introduction}
\label{sec:introduction}


Sequential recommendation predicts the next item a user will interact with from their interaction history. Generative recommenders cast this task as autoregressive decoding, which requires an item representation that a language model can emit directly. Semantic ID provides one: each catalog item is represented as a short sequence of discrete tokens obtained by quantizing its content embedding, so that semantically related items share prefixes \cite{rajput2023recommender,wang2024learnable}. This representation allows LLM-based recommenders to generate items within the same autoregressive interface used for language. Recent work first produces a natural-language reasoning trace and then predicts an item \cite{you2026r,zhu2026learning}; in SID-based methods, the trace precedes autoregressive decoding of a next-item SID \cite{liu2025onerec,hong2025generative,he2026reasoning}. A single rollout therefore spans two heterogeneous segments: the trace, for which no ground-truth target exists, and a discrete item, for which the held-out item supplies exact supervision. Once reinforcement learning is applied to the joint rollout, the central question is how final recommendation feedback should assign credit to the trace.


A common group-relative objective scores each sampled rollout through its generated SID, using exact item correctness or prefix-level matching. These rewards appropriately supervise item decoding but provide weak feedback for the intermediate tokens. Such an objective can separate two rollouts only when the reward assigns them different values, and exact match rarely does so in a large catalog: with a finite rollout group, every sampled SID may miss the held-out item \cite{liu2025onerec,zhang2026hcgrec}, leaving the reward constant within the group so that normalization produces zero SID-derived advantage. Even when SID rewards do vary, rollouts sharing a reward value receive identical SID-derived credit, and their traces are treated as equivalent. Prefix-level matching yields a graded score and therefore mitigates the constant-reward case, but it remains defined over decoded item tokens and still does not evaluate the trace itself.

We require the model to state future interests in a fixed format and run each
interest as a query against a frozen catalog retriever. Consider a rollout with
four interest lines that ends in the wrong SID. If only the second query
retrieves the held-out target, per-query evidence records the second line as the
only hit. We route positive retrieval advantage to that line, not to the other
three lines or the wrong SID. For a non-covering rollout in the same group, no
query has positive evidence, so its negative relative advantage is shared
across its valid interest lines.

The response contains a history summary and several interest queries, followed
by a valid SID. A binary rollout reward is one when any query retrieves the
held-out target within the \mbox{top-$K$}; a separate indicator retains the
outcome of every query for attribution. Generated queries and personas have
previously served as retrieval inputs
\cite{acharya2025gloss,wang2026llm}; we use retrieval during training to
score the model's own interest queries. The reward measures target retrieval
under the fixed retriever. We initialize SID--language alignment and supervised reasoning activation from \cite{he2026reasoning}, adapting the activation trace to our structured interest--SID response, and then apply group-relative RL. During rollout, a prefix trie restricts generated SIDs to valid catalog items. The objective separately normalizes three rewards within
each rollout group: a structural reward for format and history citations, a
retrieval reward for generated interests, and an exact-match SID reward. The
retrieval advantage is then routed with the per-query indicators: hitting
interest lines receive the positive signal, while the valid interest lines of
a non-covering rollout share its negative signal. It is always zero over the
final SID span. If all SID rewards in a group are zero but the retrieval
rewards differ, the model can still update the interest lines. If the retrieval rewards are also
constant, this additional signal is zero.

The generated interests also provide an auxiliary recall interface. We run them against the frozen retriever to form a recall pool. Optionally, the SIDs in this
pool define a candidate-specific trie, and the same policy performs beam search
with probabilities normalized over that trie. The interests therefore restrict
the candidate space, while the SID decoder produces the final recommendation.





We summarize our contributions as follows:
\begin{itemize}
  \item We identify a structural limit of item-level rewards in common sequential recommendation methods: a reward computed from the decoded item can become finer only within the SID span, while every token of the preceding trace still receives one rollout-level advantage. 
  \item We therefore propose a process reward on the trace itself. Interests are emitted in a fixed format and executed as queries against a frozen retriever; per-query outcomes route retrieval advantage only to the interest spans whose queries retrieve the target, leaving the final SID span under item correctness. The same queries supply candidates at inference.
  \item Empirically, the proposed method improves SID recommendation across three Amazon Reviews datasets. We also quantify how often the retrieval channel reactivates SID-inactive
  groups with constant exact-match rewards,
  and use an oracle analysis to demonstrate the potential of interest-conditioned SID decoding when the target-relevant query is selected from the generated interests.
\end{itemize}

%% file: sections/related_work.tex
\section{Related Work}
\label{sec:related-work}

\subsection{Semantic-ID Generative Recommendation}

Generative recommendation predicts item identifiers rather than scoring the
full catalog. TIGER \cite{rajput2023recommender} represents each item by a short
sequence of codes obtained by quantizing its content embedding, and generates
this Semantic ID autoregressively. Later work improves how these
identifiers are learned or augments them with collaborative signals. VQ-Rec
\cite{hou2023learning} learns vector-quantized representations for transferable
recommendation; LETTER \cite{wang2024learnable} trains the tokenizer with
recommendation supervision. EAGER \cite{wang2024eager} models semantic and
behavioral signals in separate streams. LC-Rec
\cite{zheng2024adapting} and EAGER-LLM \cite{hong2025eager} further align item
tokens with pretrained language models.

SIDs have also been used beyond standard next-item prediction. Penha et al.
\cite{penha2025semantic} study a shared SID space for search and recommendation,
while SIGMA \cite{yu2026sigma} uses hybrid item tokens for several
instruction-conditioned tasks. Zeng et al. \cite{zeng2026deepinterestgr} mine
latent interests with multimodal LLMs and use them in SID construction and
reward design. Quantization, however, does not preserve every property of the
original item embedding: Wang et al. \cite{wang2026understanding} report
substantial changes in fine-grained neighborhoods after discretization. We
retain the tokenizer and SID space of the base recommender, and instead focus
on retrieval supervision during reinforcement learning optimization.

\subsection{Reasoning for Generative Recommendation}

Recent work adds either explicit or latent reasoning to recommendation.
R$^2$ec \cite{you2026r} uses separate heads for language and recommendation,
and optimizes them jointly with reinforcement learning. OneRec-Think
\cite{liu2025onerec} grounds item tokens in text, activates in-text
reasoning, and then refines the model with a recommendation reward. GREAM
\cite{hong2025generative} combines semantic alignment, a reasoning curriculum, and
group policy optimization. SIDReasoner \cite{he2026reasoning} strengthens
SID--language alignment before optimizing reasoning trajectories with the
final recommendation outcome. OneReason
\cite{team2026onereason} organizes reasoning around item perception and
user-interest cognition, while TwiSTAR \cite{cao2026twistar} learns when slower
reasoning is worth its inference cost. Other methods perform the extra computation in latent space. S$^2$GR
\cite{guo2026s2gr} inserts a supervised thinking token before each SID token.
PauseRec \cite{he2026implicit} instead uses learned pause tokens, motivated by
the observation that free-form rationales can be sensitive to rationale
quality and to the mismatch between text and SID embeddings. These studies also
expose a limitation of explicit reasoning: readable text alone does not
establish that the trace is useful. Existing methods usually judge a trace
through its final recommendation; we additionally test whether its stated
interests can retrieve the target item.

\subsection{Process Supervision}

An exact-match SID reward is sparse because an error in an early SID token
can move all sampled trajectories to the wrong catalog branch. PROMISE
\cite{guo2026promise} trains a process reward model over intermediate SID paths
and uses it to guide beam search. HCGRec \cite{zhang2026hcgrec} addresses
all-zero rollout groups by supplying a minimal target-prefix hint and assigning
different credit to the hinted prefix and sampled suffix. SAPO
\cite{zheng2026sapo} pairs each reasoning block with one SID token and computes
step-aligned match advantages to localize SID-token errors. These methods
provide finer supervision along the SID generation path. Our method instead
verifies each generated language interest through catalog retrieval and uses
per-query hit evidence to route the auxiliary advantage to the corresponding
hitting interest line rather than the complete interest block.

\subsection{Language-Based Retrieval}

Another line of work uses generated language as the retrieval input.
GLoSS \cite{acharya2025gloss} produces textual queries from interaction
histories and retrieves items with dense semantic search. Cross-modal RAG \cite{zhu2025cross} leverages query 
decomposition for fine-grained semantic retrieval. Wang et al.
\cite{wang2026llm} generate multiple interest personas that serve as an
online retrieval source in a large-scale video platform. Such systems use
language as the recall interface, whereas SIDreasoner \cite{he2026reasoning} directly generates item
tokens.

Our method connects these two interfaces in one generative policy. The model
generates future-interest statements grounded in the interaction history,
followed by a catalog-constrained SID. We reward both the structure of the
trace and whether its interest statements retrieve the held-out item. At
inference time, the same policy can provide queries for candidate recall or
directly decode the final SID.

%% file: sections/methodology.tex
\section{Methodology}
\label{sec:methodology}

Training uses two pre-RL stages adapted from SIDReasoner
\cite{he2026reasoning}, followed by retrieval-grounded policy optimization.
\textbf{Stage 1} aligns newly added SID tokens with item language and
sequential behavior. \textbf{Stage 2} teaches the model to produce a structured
reasoning trace with a history summary and future-interest statements before
generating the final SID. \textbf{Stage 3} samples these trajectories and
applies group-relative RL. Each domain uses one fixed SID tokenizer across all
stages. Each checkpoint initializes the next stage.

Stage 3 supplies feedback at three resolutions. Exact SID match scores the
final recommendation, a rule-based trace reward verifies the structure of the
reasoning trace and its citations to the interaction history, and a frozen
dense retriever evaluates whether the generated interests retrieve the target
from the item catalog. We normalize these rewards separately and apply their
advantages to the full response, the reasoning trace, and individual interest
lines, respectively. For the retrieval channel, query-attributed span routing
sends positive advantage only to interest lines whose queries hit the target. At
inference time, the generated interests form an auxiliary recall pool that can
optionally define the trie for candidate-constrained SID decoding.

\begin{figure*}[t]
  \centering
  \includegraphics[width=\textwidth]{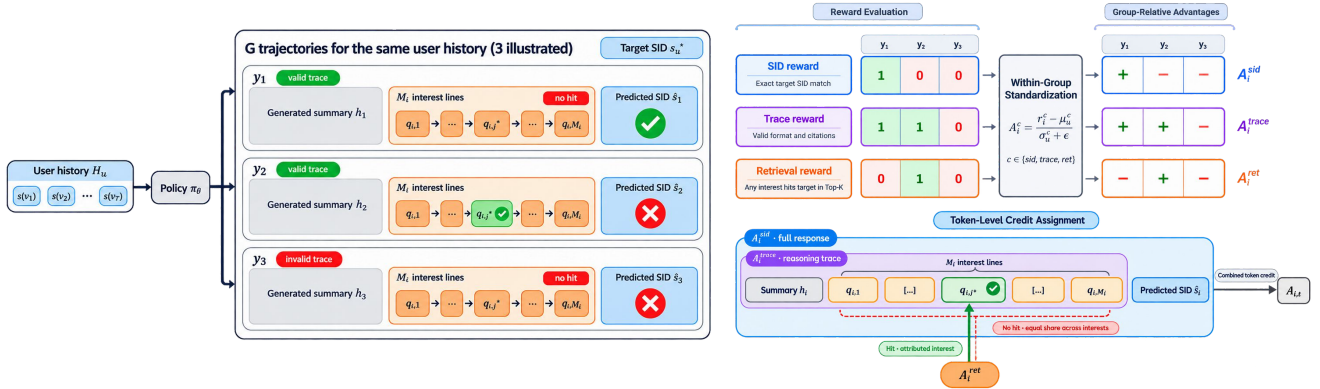}
  \caption{Grouped trajectory generation, separately standardized reward
  signals, and token-level credit assignment.}
  \label{fig:training-overview}
\end{figure*}


\subsection{Problem Definition}
\label{sec:problem-definition}

Let $I$ be the item catalog, and let
$s(v)=(s_1(v),\ldots,s_L(v))$ denote the fixed $L$-token SID of
$v\in I$. For a user history $H_u=(v_1,\ldots,v_T)$, the input $x_u$
consists of a fixed instruction followed by the concatenated history SIDs
$s(v_1),\ldots,s(v_T)$; $s_u^*$ denotes the SID of the held-out next item.
We use the domain-specific SID
tokenizers and item-to-SID mappings produced by its three-level RQ-VAE.
Accordingly, $L=3$, and each of the three codebooks contains 256 entries. The
codebooks and item-to-SID mappings remain fixed throughout all training stages.
The SID tokens are added to the language model vocabulary alongside the
existing text tokens, allowing the same autoregressive policy to generate both
the reasoning trace and the final SID.

During Stage 3, the policy samples structured trajectories. The $i$-th
trajectory contains a reasoning trace with a history summary $h_i$ and a
future-interest block $f_i$, followed by a predicted SID $\hat{s}_i$:
\begin{equation}
  y_i=[h_i;f_i;\hat{s}_i],
  \qquad
  y_i\sim\pi_\theta(\cdot\mid x_u).
  \label{eq:rollout}
\end{equation}

Let $q_{i,1},\ldots,q_{i,M_i}$ be the queries in $f_i$, and let $R_K(q)$ be
the top-$K$ items returned by a fixed catalog retriever. We assess each
trajectory along three dimensions. Its reasoning trace is valid when it
satisfies the prescribed format and history-citation rules. The SID path
succeeds when $\hat{s}_i=s_u^*$. The interest path succeeds when at least one
retrieved item has SID $s_u^*$.

\subsection{Pre-RL Initialization}
\label{sec:pre-rl-initialization}

We reuse SIDReasoner's Stage-1 alignment data
\cite{he2026reasoning}. We then construct new GPT-4o supervision for Stage 2,
adapting the reasoning format to expose retrieval queries.

\subsubsection*{Stage 1: SID Alignment}
We train the domain-specific SID tokens using the eight streams in
Table~\ref{tab:alignment-mixture}.

\begin{table}[t]
  \caption{Stage-1 SID alignment streams.}
  \label{tab:alignment-mixture}
  \small
  \setlength{\tabcolsep}{3pt}
  \begin{tabular}{@{}ll@{}}
    \toprule
    Training stream & Purpose and loss \\
    \midrule
    Title $\leftrightarrow$ SID translation
      & Bidirectional grounding (SFT) \\
    SID history $\rightarrow$ next SID
      & SID-space recommendation (SFT) \\
    SID history $\rightarrow$ next-item title
      & SID-to-title recommendation (SFT) \\
    Title history $\rightarrow$ next-item title
      & Title-space recommendation (SFT) \\
    Title history $\rightarrow$ next SID
      & Title-to-SID recommendation (SFT) \\
    Item-level SID--text interleaving
      & Item grounding (full LM) \\
    Sequence-level SID--text interleaving
      & Sequence grounding (full LM) \\
    General reasoning
      & General-ability retention (SFT) \\
    \bottomrule
  \end{tabular}
\end{table}

The first five rows and general reasoning use chat-formatted, completion-only
SFT with loss on the assistant response. The two interleaving streams use
full-sequence language modeling to place SIDs in item- and sequence-level text.
All streams jointly update the model, including the newly added SID-token
embeddings.

\subsubsection*{Stage 2: Structured Reasoning Activation}
Each completion target has the following form:
\begin{center}
  \footnotesize
  \begin{tabular}{@{}l@{}}
    \texttt{<think>} \\
    \quad\texttt{<history\_summary>} \\
    \qquad\texttt{- SID(s) => factual observation} \\
    \qquad\texttt{...} \\
    \quad\texttt{</history\_summary>} \\
    \quad\texttt{<future\_interests>} \\
    \qquad\texttt{- SID(s) => }$q_{i,1}$ \\
    \qquad\texttt{...} \\
    \quad\texttt{</future\_interests>} \\
    \texttt{</think>} \\
    \texttt{target SID}
  \end{tabular}
\end{center}
The ellipses denote additional entries in each block.

\paragraph{Trace fields.}
A summary line pairs cited history SIDs with a factual observation. A
future-interest line states a plausible next interest grounded in its cited
history SIDs. For line $j$, $q_{i,j}$ is the text after \texttt{=>} and is the
only part sent to the retriever. The tags locate its line-level token mask in
Stage 3.

\paragraph{Stage-2 data.}
Following \cite{he2026reasoning}, we use an LLM teacher to generate the SFT traces. For
each training instance, the teacher receives the interaction history and item
metadata together with the held-out next-item metadata, and generates a
reasoning trace in the schema above. We pair the trace with the target SID to
form the assistant completion. Stage 2 uses completion-only SFT for one epoch,
and its checkpoint initializes Stage 3.

\subsection{Retrieval-Grounded Policy Optimization}
\label{sec:policy-optimization}

Stage 3 assigns credit at two levels. Across the $G$ trajectories sampled for
the same user history, separately normalized SID, trace, and retrieval rewards
determine the relative credit of each trajectory. Within a trajectory,
per-query retrieval evidence attributes retrieval credit to individual
interest lines. We combine both levels into token advantages: SID credit
applies to the full response, trace credit to the reasoning trace, and
retrieval credit only to attributed interest lines.

\subsubsection{Catalog-Constrained Rollouts}
Each trajectory first generates the two text blocks and the closing
\texttt{</think>} marker, then samples its $L$-token SID autoregressively. We
build a prefix trie from the catalog SIDs so every sampled sequence completes
to a catalog SID. Let $r$ denote the generated text trace, and suppress the
trajectory index in this subsection. At SID position $l$, let
$\hat{s}_{<l}$ be the current prefix, let
$\mathcal{V}(\hat{s}_{<l})$ be its valid next tokens, and let
$c_l=[x_u;r;\hat{s}_{<l}]$ be the full autoregressive context. The sampling
distribution is
\begin{equation}
  \pi_\theta^{\mathrm{trie}}(a\mid c_l)
  =
  \frac{
    \exp z_\theta(a\mid c_l)
  }{
    \sum_{a'\in \mathcal{V}(\hat{s}_{<l})}
    \exp z_\theta(a'\mid c_l)
  },
  \quad a\in \mathcal{V}(\hat{s}_{<l}).
  \label{eq:trie-policy}
\end{equation}
Tokens outside $\mathcal{V}(\hat{s}_{<l})$ have probability zero. This is the
standard language-model softmax restricted to the outgoing edges of the
current trie node.

\subsubsection{Reward Signals}
Each trajectory produces three binary rollout-level rewards. The retrieval
channel additionally retains per-query hit indicators for subsequent
within-trajectory attribution.

\paragraph{SID exact-match reward.}
The SID reward records exact agreement between the sampled and target SIDs:
$r_i^{\mathrm{sid}}=\mathbf{1}[\hat{s}_i=s_u^*]$.

\paragraph{Trace-validity reward.}
The trace reward is one when the response contains exactly one summary block
followed by one interest block and every cited SID occurs in the input history.
We write this reward as
$r_i^{\mathrm{trace}}
=\mathbf{1}[\text{valid trace}]$.

\paragraph{Interest-retrieval reward.}
A frozen catalog retriever ranks items for each parsed interest query. For a
query $q$, let $\operatorname{rank}(s\mid q)$ be the best rank of any catalog
item whose SID is $s$, or $+\infty$ if no such item is retrieved. For cutoff
$K$, we define the query-level evidence and rollout-level retrieval reward as
\begin{equation}
  \begin{aligned}
    z_{i,j}^{\mathrm{ret}}
    &=
    \mathbf{1}[\text{interest block is valid}]\,
    \mathbf{1}
    [\operatorname{rank}(s_u^*\mid q_{i,j})\le K],\\
    r_i^{\mathrm{ret}}
    &=
    \mathbf{1}\!\left[
      \sum_{j=1}^{M_i}z_{i,j}^{\mathrm{ret}}>0
    \right].
  \end{aligned}
  \label{eq:retrieval-evidence}
\end{equation}
Thus, $r_i^{\mathrm{ret}}=1$ if at least one query retrieves the target within
the top-$K$; it is zero when no valid query is parsed. We retain each
$z_{i,j}^{\mathrm{ret}}$ for line-level routing. We use a top-$K$ event rather
than the raw retrieval score because the desired behavior is candidate
coverage, not score calibration.

\subsubsection{Across-Trajectory Credit}
The rollout rewards determine which trajectories receive positive or negative
relative credit within the group sampled for one user history.
For each reward
$c\in\{\mathrm{sid},\mathrm{trace},\mathrm{ret}\}$, we compute
\begin{equation}
  \begin{aligned}
    \mu_u^c
    &=
    \frac{1}{G}\sum_{i=1}^{G}r_i^c,\\
    \sigma_u^c
    &=
    \sqrt{
      \frac{1}{G-1}
      \sum_{i=1}^{G}(r_i^c-\mu_u^c)^2
    },\\
    A_i^c
    &=
    \frac{r_i^c-\mu_u^c}{\sigma_u^c+\epsilon}.
  \end{aligned}
  \label{eq:channel-advantages}
\end{equation}
The denominator $G-1$ matches the sample standard deviation used in the
implementation. Following group-relative optimization
\cite{shao2024deepseekmath}, we standardize rewards within each rollout group.
We apply this operation separately to the SID, trace, and retrieval rewards.
Whenever a reward varies within a group, its normalized advantages have a
comparable scale before weighting. If a reward is constant within a group, it
contributes zero advantage.

\subsubsection{Token-Level Credit Assignment}
The three rewards assign credit to different parts of a trajectory. As the
final outcome signal, the SID advantage applies to the full response; the
trace advantage applies to the reasoning trace; and the retrieval advantage is
attributed to interest lines using query-level retrieval evidence.

\paragraph{Query-level routing.}
The trajectory-level advantage $A_i^{\mathrm{ret}}$ determines the sign and
relative magnitude of retrieval credit, while $w_{i,j}$ determines which
interest lines receive it. For a retrieval-positive trajectory, the binary hit
indicators divide the credit equally among the hitting interest lines. For a
retrieval-negative trajectory in a mixed group, no query provides positive
evidence, so the negative credit is divided equally among all parsed interest
lines. We define
\begin{equation}
  w_{i,j}
  =
  \begin{cases}
    \dfrac{z_{i,j}^{\mathrm{ret}}}
          {\sum_{\ell=1}^{M_i}z_{i,\ell}^{\mathrm{ret}}},
      & r_i^{\mathrm{ret}}=1,\\[6pt]
    \dfrac{1}{M_i},
      & r_i^{\mathrm{ret}}=0,\ M_i>0,\\[4pt]
    0,
      & M_i=0.
  \end{cases}
  \label{eq:query-routing}
\end{equation}
For $M_i>0$, the line weights sum to one. Positive trajectories select the
hitting interest lines, whereas negative trajectories select all parsed
interest lines.

\paragraph{Token-level composition.}
For token $t$ in trajectory $i$, $m_{i,j,t}^{\mathrm{line}}$ indicates whether
it belongs to the $j$-th interest line, and $m_{i,t}^{\mathrm{trace}}$
indicates whether it belongs to the reasoning trace. These masks localize the
retrieval and trace advantages to their corresponding spans. If an interest
line cannot be located unambiguously, its mask is zero.

The retriever evaluates only the query text, while the routed retrieval
advantage is applied to the corresponding interest line. Combining these
credit scopes gives the token-level advantage:
\begin{equation}
  \begin{aligned}
    A_{i,t}
    ={}
    A_i^{\mathrm{sid}}
    +
    \lambda_{\mathrm{trace}}\,
    m_{i,t}^{\mathrm{trace}}A_i^{\mathrm{trace}}+
    \lambda_{\mathrm{ret}}\,
    \left(
    \sum_{j=1}^{M_i}
    w_{i,j}m_{i,j,t}^{\mathrm{line}}
    \right)
    A_i^{\mathrm{ret}}.
  \end{aligned}
  \label{eq:token-advantage}
\end{equation}
The retrieval advantage is divided across the attributed interest lines
through $w_{i,j}$ and applied uniformly to their tokens. Positions in the final
SID span receive only $A_i^{\mathrm{sid}}$.

\begin{table}[t]
  \caption{Reward channels and token-level credit scopes.}
  \label{tab:reward-routing}
  \small
  \setlength{\tabcolsep}{3pt}
  \begin{tabular}{@{}lll@{}}
    \toprule
    Reward & Criterion & Credit scope \\
    \midrule
    SID & Exact match & Full response \\
    Trace & Valid structure & Reasoning trace \\
    Retrieval & Target coverage & Attributed interest lines \\
    \bottomrule
  \end{tabular}
\end{table}

Table~\ref{tab:reward-routing} summarizes the resulting token-level credit
scopes.

\subsubsection{SID-Inactive Groups}
A rollout group is SID-inactive when its SID exact-match rewards are constant,
which gives $A_i^{\mathrm{sid}}=0$ for every trajectory. The common all-zero
case occurs when none of the sampled SIDs matches the target. If retrieval
rewards vary within such a group, the retrieval channel can still assign
nonzero credit to attributed interest lines and distinguish trajectories whose
queries retrieve the target from those whose queries do not.

\subsubsection{Policy Optimization}
Let $\rho_{i,t}$ be the current-to-old probability ratio for response token
$t$. Both its numerator and denominator use standard language-model
probabilities on reasoning tokens and the trie-restricted distribution in
Eq.~\eqref{eq:trie-policy} on final SID tokens, thereby matching the rollout
action space. Let $d_{i,t}^{\mathrm{KL}}$ denote the token-level KL divergence
from the frozen pre-RL policy, evaluated over the same action space. With
$N=\sum_{i=1}^{G}|y_i|$ and coefficient $\beta_{\mathrm{KL}}\ge0$, we optimize
\begin{equation}
  \mathcal{L}_{\mathrm{actor}}
  =
  -\frac{1}{N}
  \sum_{i=1}^{G}
  \sum_{t=1}^{|y_i|}
  \left[
    \ell_{\mathrm{PPO}}(\rho_{i,t},A_{i,t})
    -
    \beta_{\mathrm{KL}}d_{i,t}^{\mathrm{KL}}
  \right].
  \label{eq:actor-loss}
\end{equation}
Here $\ell_{\mathrm{PPO}}$ is the dual-clip PPO surrogate
\cite{ye2020mastering}.

\subsection{SID Decoding with Auxiliary Interest Recall}
\label{sec:auxiliary-interest-recall}

\paragraph{SID decoding.}
Our primary inference mechanism is SID decoding. Given a user history, the
policy first generates a structured trace. Conditioned on the history and this
trace, catalog-constrained beam search returns a ranked list
$C_{\mathrm{sid}}$ of valid SID candidates, which serves as the
direct recommendation output.

\paragraph{Auxiliary interest recall.}
The retrieval-supervised interests additionally provide an auxiliary recall
interface. We issue each generated interest query to the frozen catalog
retriever used during training, merge the per-query top-$K$ results, and remove
duplicate SIDs to obtain a recall pool $C_{\mathrm{ret}}$. This pool exposes
catalog coverage captured by the generated interests.

\paragraph{Optional candidate-constrained SID decoding.}
The recall pool can also define a candidate-specific SID trie. We build this
trie from the SIDs in $C_{\mathrm{ret}}$ and, conditioned on the same history
and trace, perform beam search with probabilities normalized over the candidate
trie. This returns a ranked list $C_{\mathrm{rcd}}$ through
candidate-constrained generation rather than reranking candidates by their
full-catalog SID probabilities.

%% file: sections/experiments_arxiv.tex
\section{Experiments}
\label{sec:experiments}

\subsection{Experimental Setup}
\label{sec:experimental-setup}

  \begin{table}[t]
    \caption{Dataset statistics after preprocessing.}
    \label{tab:dataset-statistics}
    \centering
    \setlength{\tabcolsep}{1.6pt}
    \begin{tabular}{@{}lrrrrrr@{}}
      \toprule
      Category & Users & Items & Train & Val. & Test & Avg.\ len. \\
      \midrule
      Video Games & 6,142 & 3,858 & 49,133 & 6,142 & 6,142 & 6.45 \\
      Office & 4,866 & 3,459 & 38,924 & 4,866 & 4,866 & 5.97 \\
      Industrial & 4,533 & 3,686 & 36,259 & 4,532 & 4,533 & 5.96 \\
      \bottomrule
    \end{tabular}
  \end{table}

\begin{table*}[t]
  \caption{Overall recommendation performance. The best result in each column
  is shown in bold. The SID-plus-trace control and our method report means over
  three training seeds; their maximum standard deviations across all columns
  are 0.0053 and 0.0039, respectively. An asterisk (*) indicates a statistically
  significant improvement over the SID-plus-trace-validity control under a
  paired two-sided $t$-test over test users ($p<0.05$).}
  \label{tab:main-results}
  \centering
  \scriptsize
  \setlength{\tabcolsep}{3.5pt}
  \begin{tabular}{@{}lcccccccccccc@{}}
    \toprule
    & \multicolumn{4}{c}{Video Games}
    & \multicolumn{4}{c}{Office Products}
    & \multicolumn{4}{c}{Industrial and Scientific} \\
    \cmidrule(lr){2-5}\cmidrule(lr){6-9}\cmidrule(l){10-13}
    Method
    & Recall@5 & NDCG@5 & Recall@10 & NDCG@10
    & Recall@5 & NDCG@5 & Recall@10 & NDCG@10
    & Recall@5 & NDCG@5 & Recall@10 & NDCG@10 \\
    \midrule
    \rowcolor{controlrow}
    \multicolumn{13}{@{}l}{\textit{Traditional sequential recommenders}} \\
    GRU4Rec
      & 0.0329 & 0.0219 & 0.0599 & 0.0305
      & 0.0682 & 0.0480 & 0.0974 & 0.0574
      & 0.0788 & 0.0578 & 0.1030 & 0.0649 \\
    SASRec
      & 0.0501 & 0.0345 & 0.0723 & 0.0416
      & 0.1019 & 0.0824 & 0.1167 & 0.0871
      & 0.0807 & 0.0647 & 0.0964 & 0.0697 \\
    Caser
      & 0.0376 & 0.0241 & 0.0659 & 0.0332
      & 0.0880 & 0.0663 & 0.1114 & 0.0738
      & 0.0664 & 0.0528 & 0.0852 & 0.0588 \\
    \rowcolor{controlrow}
    \multicolumn{13}{@{}l}{\textit{Generative recommenders without reasoning}} \\
    TIGER
      & 0.0489 & 0.0300 & 0.0763 & 0.0402
      & 0.1270 & 0.1037 & 0.1429 & 0.1121
      & 0.1003 & 0.0823 & 0.1325 & 0.0924 \\
    HSTU
      & 0.0539 & 0.0396 & 0.0746 & 0.0462
      & 0.1204 & 0.1069 & 0.1323 & 0.1107
      & 0.1008 & 0.0898 & 0.1138 & 0.0940 \\
    LETTER
      & 0.0445 & 0.0294 & 0.0709 & 0.0378
      & 0.1315 & 0.1074 & 0.1520 & 0.1139
      & 0.1080 & 0.0850 & 0.1389 & 0.0950 \\
    LCRec
      & 0.0441 & 0.0274 & 0.0876 & 0.0412
      & 0.0964 & 0.0699 & 0.1487 & 0.0867
      & 0.0805 & 0.0520 & 0.1330 & 0.0687 \\
    \rowcolor{controlrow}
    \multicolumn{13}{@{}l}{\textit{Reasoning-based recommenders}} \\
    ReaRec
      & 0.0568 & 0.0381 & 0.0843 & 0.0470
      & 0.1173 & 0.0988 & 0.1385 & 0.1057
      & 0.0973 & 0.0796 & 0.1205 & 0.0870 \\
    R$^2$ec
      & 0.0655 & 0.0399 & 0.0931 & 0.0525
      & 0.1147 & 0.0894 & 0.1486 & 0.1004
      & 0.0880 & 0.0774 & 0.1253 & 0.0774 \\
    SIDReasoner
      & 0.0710 & 0.0460 & 0.1031 & 0.0563
      & 0.1373 & 0.1119 & 0.1648 & 0.1208
      & 0.1109 & 0.0905 & 0.1438 & 0.1010 \\
    \midrule
    \rowcolor{controlrow}
    \multicolumn{13}{@{}l}{\textit{Controlled Stage-3 RL ablations}} \\
    Stage-2 SFT checkpoint (no RL)
      & 0.0591 & 0.0407 & 0.0858 & 0.0493
      & 0.1297 & 0.1074 & 0.1506 & 0.1142
      & 0.1125 & 0.0893 & 0.1370 & 0.0972 \\
    Stage-3 RL: SID exact-match reward
      & 0.0583 & 0.0399 & 0.0861 & 0.0489
      & 0.1281 & 0.1046 & 0.1505 & 0.1119
      & 0.1098 & 0.0856 & 0.1371 & 0.0944 \\
    + Trace-validity reward
      & 0.0591 & 0.0404 & 0.0856 & 0.0493
      & 0.1272 & 0.1051 & 0.1512 & 0.1124
      & 0.1116 & 0.0868 & 0.1387 & 0.0953 \\
    \rowcolor{oursrow}
    \textbf{+ Query-attributed retrieval reward (Ours)}
      & \textbf{0.0879}\textsuperscript{*}
      & \textbf{0.0694}\textsuperscript{*}
      & \textbf{0.1195}\textsuperscript{*}
      & \textbf{0.0796}\textsuperscript{*}
      & \textbf{0.1515}
      & \textbf{0.1251}
      & \textbf{0.1765}\textsuperscript{*}
      & \textbf{0.1332}\textsuperscript{*}
      & \textbf{0.1291}
      & \textbf{0.1043}
      & \textbf{0.1602}\textsuperscript{*}
      & \textbf{0.1144}\textsuperscript{*} \\
    \rowcolor{controlrow}
    Absolute $\Delta$ vs.\ SID+trace
      & +0.0288 & +0.0290 & +0.0339 & +0.0303
      & +0.0243 & +0.0200 & +0.0253 & +0.0208
      & +0.0175 & +0.0175 & +0.0215 & +0.0191 \\
    \bottomrule
  \end{tabular}
\end{table*}

\paragraph{Datasets and protocol.}
  We use the Video Games, Office Products, and Industrial and Scientific
  categories of the Amazon Reviews corpus \cite{ni2019justifying}. Following the
processed data used by SIDReasoner \cite{he2026reasoning}, we retain
interactions from October 2016 through November 2018, apply 5-core filtering,
sort each user's interactions chronologically, and truncate histories to the
most recent 10 items. The last interaction of each user is held out for test,
the second-to-last for validation, and earlier interactions form sliding-window
training pairs. Table~\ref{tab:dataset-statistics} summarizes the resulting
data.

\newcommand{\resultplaceholder}{\mbox{\phantom{0.0000}}}

\paragraph{Evaluation metrics.}
We use the SIDReasoner evaluator. Table~\ref{tab:main-results} includes its
reported baselines \cite{he2026reasoning}. For each test instance, the
evaluator performs catalog-constrained beam search using a trie containing all
catalog SIDs and returns a ranked beam of ten valid SIDs. Evaluation uses no
sampled negatives. We report Recall and NDCG at cutoffs 5 and 10.
For interest retrieval, Interest Recall@50 (I-R@50) is the fraction of
examples for which at least one generated interest retrieves the held-out
target within the frozen retriever's top 50 results. Our model generates three
to four interests per instance, with an average of 3.8.

\paragraph{Baselines.}
We compare against three families of recommenders:
\begin{itemize}
  \item \textbf{Traditional sequential recommenders:} GRU4Rec
  \cite{hidasi2015session}, SASRec \cite{kang2018self}, and Caser
  \cite{tang2018personalized}, which rank items directly from the interaction
  sequence.
  \item \textbf{Generative recommenders without reasoning:} TIGER
  \cite{rajput2023recommender}, HSTU \cite{zhai2024actions}, LETTER
  \cite{wang2024learnable}, and LCRec \cite{zheng2024adapting}.
  \item \textbf{Reasoning-based methods:}\newline
  ReaRec
  \cite{tang2026think}, R$^2$ec \cite{you2026r}, and SIDReasoner
  \cite{he2026reasoning}.
\end{itemize}

\paragraph{Implementation details.}
Based on Qwen3-1.7B \cite{yang2025qwen3}, we fine-tune all parameters in every stage, paired with the same domain-specific three-level RQ-VAE tokenizer. Stage 1 uses AdamW for up to five epochs with a learning rate of
$2\times10^{-5}$. Stage 2 uses AdamW for one epoch with a learning rate of
$1\times10^{-5}$.

Stage 3 samples 16 trajectories per prompt with temperature 1.0 and
top-$p=1.0$. We use a global batch of 256 prompts, maximum prompt and response
lengths of 1,024 tokens each, an actor learning rate of $7\times10^{-7}$, and
dual-clip PPO with a clip ratio of 0.2 and dual-clip constant 3.0. Training
runs for at most 10 epochs, with checkpoint selection by validation NDCG@10.
The default configuration omits KL regularization because a Video Games
sensitivity check found no material effect on recommendation performance.

The trace-validity and retrieval rewards are each weighted by $0.1$.
Each catalog document contains labeled title, optional brand, and description
lines. Their Qwen3-Embedding-4B embeddings \cite{zhang2025qwen3} are precomputed and frozen; each
generated interest is encoded online and matched by exact cosine search. We
use $K=50$ by default and report sensitivity across
$K\in\{10,20,50,100\}$.

All hyperparameters, checkpoint selections, and reward cutoffs are selected
using validation data. Across all three domains, the SID-plus-trace control and
our method report the mean and standard deviation over three Stage-3 seeds to
quantify training variability.

\begin{figure*}[t]
  \centering
  \includegraphics[width=0.85\textwidth]{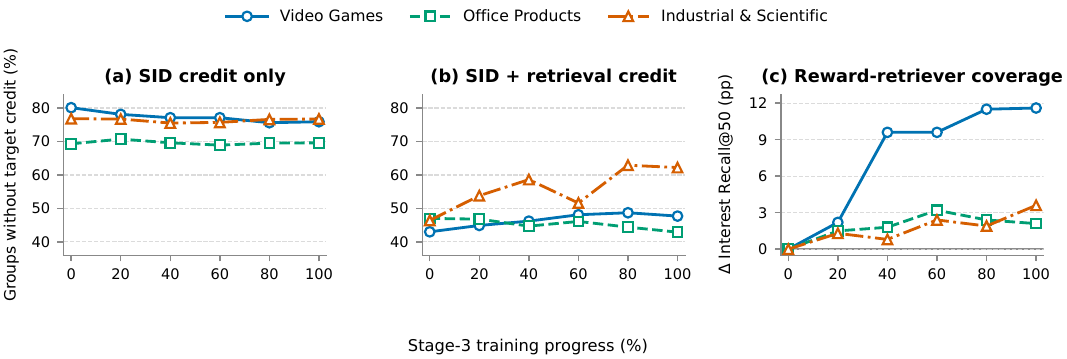}
  \caption{Stage-3 no-signal and retrieval-reactivation dynamics.
  (a) SID no-signal rate: the percentage of groups with constant binary SID
  exact-match rewards. (b) Joint no-signal rate: the percentage with both
  constant SID and constant retrieval rewards. (c) Validation Interest
  Recall@50 difference, in percentage points, between our method and the
  matched SID-plus-trace control under the reward retriever.}
  \Description{Three line charts over normalized Stage-3 training progress
  compare three Amazon domains. The first shows groups with constant SID
  rewards, the second shows groups that remain without SID or retrieval
  contrast, and the third shows validation Interest Recall at 50 gains under
  the reward retriever.}
  \label{fig:reactivation-dynamics}
\end{figure*}

\subsection{Overall Recommendation Performance}
\label{sec:overall-performance}

RQ1 asks whether retrieval-grounded credit improves direct full-catalog SID
recommendation. Table~\ref{tab:main-results} compares Stage-2 SFT with three
incremental RL variants. The matched no-retrieval control is the
SID-exact-match-plus-trace-validity variant. All three RL variants start from
the same Stage-2 checkpoint and use the same data order, training budget, SID
action space, and KL configuration.

Retrieval-grounded credit consistently improves direct SID recommendation.
Relative to the matched no-retrieval control, our method improves
Recall@10 by 15.5\%--39.6\% and NDCG@10 by 18.5\%--61.5\% across the three
domains. The gains are largest on Video Games and more moderate on Office
Products and Industrial and Scientific. R$^2$ec shows a similar cross-domain
pattern, suggesting that the benefit of explicit reasoning may depend on how
well item metadata aligns with semantic knowledge available to the language
model. Improvements also hold at cutoff 5. In contrast, SID exact-match and
trace-validity optimization remain close to the Stage-2 checkpoint. This
pattern is consistent with the gain not being explained by RL or structural
supervision alone.
Section~\ref{sec:credit-mechanism} examines the additional learning signal
supplied by retrieval.

Across all baselines in Table~\ref{tab:main-results}, our method achieves the
best result on all 12 metrics.

\subsection{Retrieval Credit and Routing}
\label{sec:credit-mechanism}

RQ2 examines two questions: how often retrieval feedback supplies a nonzero
advantage when the SID reward is constant, and whether assigning this
advantage to queries that produce a retrieval hit improves recommendation over
coarser credit scopes. We analyze the SID and retrieval channels because both
provide target-dependent feedback. The trace-validity reward is nearly always
constant within a group and therefore usually produces zero group-relative
advantage.

Let $\mathcal{G}_t$ be the rollout groups in training window $t$, with $G$
trajectories in each group. A reward channel produces zero group-relative
advantage when its rewards are constant across the group. For
$c\in\{\mathrm{sid},\mathrm{ret}\}$, let $u_g^c$ indicate this no-signal
condition for channel $c$ in group $g$. The no-signal and conditional
reactivation rates are
\begin{equation}
  \begin{aligned}
    u_g^c
    &=
    \mathbf{1}\!\left[
      \max_i r_{g,i}^{c}
      =
      \min_i r_{g,i}^{c}
    \right],\\
    \mathcal{Z}_{\mathrm{sid}}(t)
    &=
    \frac{1}{|\mathcal{G}_t|}
    \sum_{g\in\mathcal{G}_t}
    u_g^{\mathrm{sid}},\\
    \mathcal{Z}_{\mathrm{sid+ret}}(t)
    &=
    \frac{1}{|\mathcal{G}_t|}
    \sum_{g\in\mathcal{G}_t}
    u_g^{\mathrm{sid}}u_g^{\mathrm{ret}},\\
    \mathcal{R}_{\mathrm{react}}(t)
    &=
    1-
    \frac{\mathcal{Z}_{\mathrm{sid+ret}}(t)}
         {\mathcal{Z}_{\mathrm{sid}}(t)},
    \qquad \mathcal{Z}_{\mathrm{sid}}(t)>0.
  \end{aligned}
  \label{eq:no-signal-rates}
\end{equation}
$\mathcal{Z}_{\mathrm{sid}}$ is the fraction of groups with no SID-derived
advantage, while $\mathcal{Z}_{\mathrm{sid+ret}}$ is the fraction for which
both SID and retrieval produce zero group-relative advantage. Their relative
reduction, $\mathcal{R}_{\mathrm{react}}$, is the fraction of SID-inactive
groups reactivated by retrieval. Because both rewards are binary, a no-signal
group can be all zero or all one; both cases produce zero group-relative
advantage. The reported rates therefore characterize optimization signal, not
exact-match sparsity alone. 

Figure~\ref{fig:reactivation-dynamics} plots these quantities over training.
Panels (a) and (b) aggregate all logged groups from the same run of our method.
Each point uses a 5\%-wide window centered at the displayed progress, computed
as completed Stage-3 updates divided by total updates. Panel (c) evaluates the
corresponding checkpoints on a fixed validation set and reports the Interest
Recall@50 difference, in percentage points (pp), between our method and the
matched SID-plus-trace control.

\begin{figure*}[!t]
  \centering
    \includegraphics[width=0.85\textwidth]{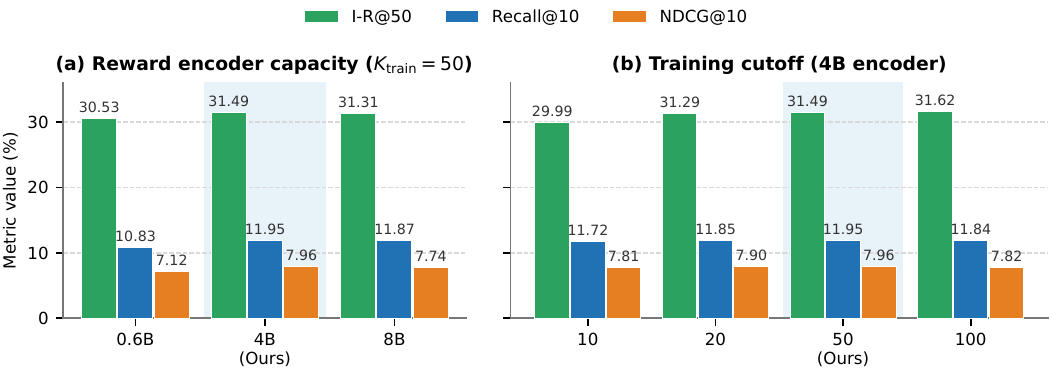}
  \caption{Reward-design sensitivity on Video Games. Bars report absolute
  Interest Recall@50, Recall@10, and NDCG@10 in percent. Panel (a) varies
  reward-encoder capacity at a fixed $K_{\mathrm{train}}=50$; panel (b) varies
  the training cutoff with the 4B encoder. Shaded groups mark our selected
  4B encoder and $K_{\mathrm{train}}=50$ cutoff.}
  \Description{Two grouped bar charts show absolute Interest Recall at 50,
  Recall at 10, and NDCG at 10. The 0.6B encoder performs worse than 4B, while
  8B is similar. Recommendation metrics vary little across training cutoffs
  from 10 to 100.}
  \label{fig:reward-sensitivity}
\end{figure*}

\paragraph{Training dynamics.}
The SID channel produces zero group-relative advantage in
$69.1\%$--$80.8\%$ of groups across the displayed domains and training
windows. By construction,
$\mathcal{Z}_{\mathrm{sid+ret}}\le\mathcal{Z}_{\mathrm{sid}}$; empirically,
the gap never falls below $27.8$ pp on Games, $22.5$ pp on Office, or
$14.7$ pp on Industrial. In the last displayed training window,
$\mathcal{R}_{\mathrm{react}}$ is $37.8\%$ on Games, $37.7\%$ on Office, and
$19.6\%$ on Industrial.

Panel (c) evaluates the generated interests with the same retriever used to
compute the training reward. At the final displayed checkpoint, our method
improves Interest Recall@50 over the SID-plus-trace control by $11.0$ pp on
Video Games, $2.2$ pp on Office Products, and $3.6$ pp on Industrial and
Scientific. The retrieval reward has its largest measured effect on Video
Games. Because training and evaluation use the same retriever, this panel
measures learning progress rather than independent generalization.

\paragraph{Credit-routing ablations.}
Table~\ref{tab:routing-ablation} compares five Stage-3 variants in all three
domains. Each result averages three seeds. Within a domain, the variants share
the Stage-2 checkpoint and training configuration, and all retain the SID
exact-match and trace-validity rewards. The SID-prefix variant adds the partial
reward used by SIDReasoner \cite{he2026reasoning}. Let $m_i$ be the length of the longest
correct prefix between the predicted and target SIDs. The reward is
\begin{equation}
  r_i^{\mathrm{prefix}}
  =
  \frac{1}{2^{L-m_i}},
  \label{eq:prefix-control}
\end{equation}
which increases toward one as more leading SID tokens match and equals one
for an exact match. After within-group normalization, its advantage is applied
to the full response, matching the scope of the exact-match SID advantage.

The remaining variants use the same retrieval reward and differ only in where
its advantage is applied. Broadcast applies it to the full response, while
block routing confines it to the future-interest block. Hit-aware routing
further uses the indicators in Eq.~\eqref{eq:query-routing}: on a positive
rollout, only queries that retrieve the target receive retrieval credit. These
controls separate the effect of denser SID supervision from the effects of
retrieval feedback and query-level attribution.

\begin{table}[t]
  \caption{Credit-routing ablations across three domains, averaged over three
  seeds.}
  \label{tab:routing-ablation}
  \centering
  \resizebox{0.95\columnwidth}{!}{%
  \begin{tabular}{lcccccc}
    \toprule
    & \multicolumn{2}{c}{Games}
    & \multicolumn{2}{c}{Office}
    & \multicolumn{2}{c}{Industrial} \\
    \cmidrule(lr){2-3}\cmidrule(lr){4-5}\cmidrule(l){6-7}
    Variant
      & R@10 & N@10
      & R@10 & N@10
      & R@10 & N@10 \\
    \midrule
    SID exact + trace
      & 0.0856 & 0.0493
      & 0.1512 & 0.1124
      & 0.1387 & 0.0953 \\
    SID-prefix
      & 0.0996 & 0.0569
      & 0.1587 & 0.1209
      & 0.1498 & 0.1001 \\
    Full broadcast
      & 0.0953 & 0.0559
      & 0.1644 & 0.1171
      & 0.1442 & 0.1039 \\
    Interest block
      & 0.1073 & 0.0610
      & 0.1662 & 0.1256
      & 0.1541 & 0.1082 \\
    \rowcolor{oursrow}
    \textbf{Hit query (Ours)}
      & \textbf{0.1195} & \textbf{0.0796}
      & \textbf{0.1765} & \textbf{0.1332}
      & \textbf{0.1602} & \textbf{0.1144} \\
    \bottomrule
  \end{tabular}%
  }
\end{table}

All four auxiliary variants improve on the SID-plus-trace control for every
domain and metric. Routing retrieval advantage to the interest block
consistently beats broadcasting it to the full response. Hit-query routing
performs best on all six metrics. Compared with block routing, its Recall@10
and NDCG@10 gains are 11.4\% and 30.5\% on Games, 6.2\% and 6.1\% on
Office, and 4.0\% and 5.7\% on Industrial. This
consistent ordering supports assigning retrieval credit to the queries
responsible for target coverage.


\subsection{Reward Validity and Interest Utility}
\label{sec:reward-validity}

\begin{table}[t]
  \caption{Overlap between interest retrieval at 50 and direct SID decoding at
  10 for our model on Video Games.}
  \label{tab:interest-sid-alignment}
  \centering
  \resizebox{0.95\columnwidth}{!}{%
  \begin{tabular}{lrrr}
    \toprule
    & SID hit@10 & SID miss@10 & Total \\
    \midrule
    Interest hit@50
      & 533 (8.68\%) & 1,401 (22.81\%) & 1,934 (31.49\%) \\
    Interest miss@50
      & 201 (3.27\%) & 4,007 (65.24\%) & 4,208 (68.51\%) \\
    \midrule
    Total
      & 734 (11.95\%) & 5,408 (88.05\%) & 6,142 \\
    \bottomrule
  \end{tabular}%
  }
\end{table}

\paragraph{Interest--SID alignment.}
Table~\ref{tab:interest-sid-alignment} cross-tabulates Interest Recall@50 and
SID Recall@10 on Video Games. SID Recall@10 is 27.56\% when the interests
retrieve the target and 4.78\% otherwise. This 22.78 percentage points gap has a 95\%
bootstrap CI of [20.70, 24.88]. The odds ratio is 7.58
(95\% bootstrap CI [6.40, 9.06]).

The overlap is partial. Among targets absent from the SID beam, 25.91\% are
still retrieved by a generated interest. Retrieval success thus tracks SID
accuracy while exposing target coverage that the final SID outcome does not
capture.

\begin{figure*}[t]
  \centering
  \includegraphics[width=0.9\textwidth]{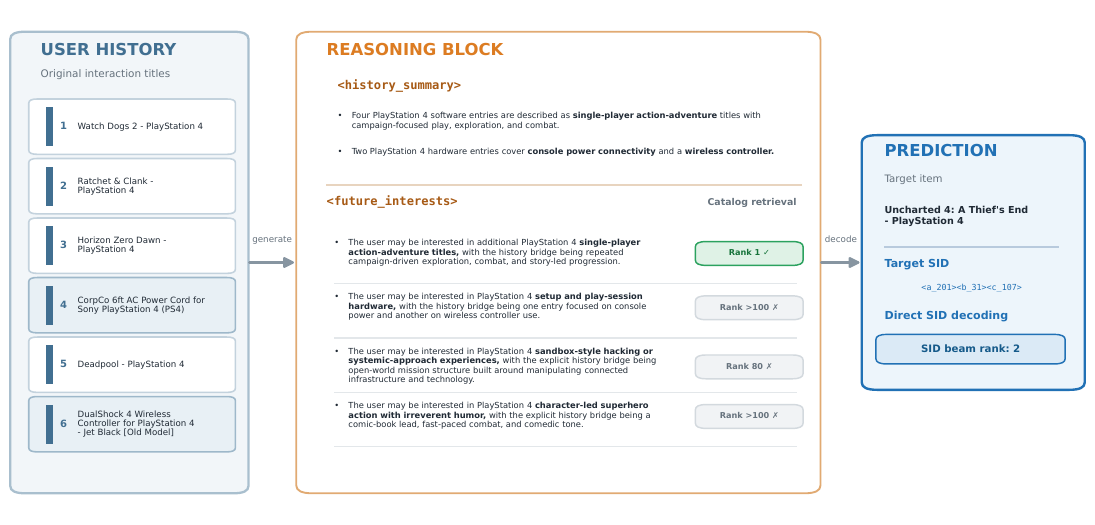}
  \caption{A Video Games case where both paths find the target. The
  action-adventure interest retrieves it at rank 1, while the other interests
  miss at the training cutoff. The SID beam ranks it second. History SIDs are
  shown as item titles, line-level citations are omitted, and bold text
  highlights each statement's key phrase.}
  \Description{The figure follows one example from six user-history items
  through a generated history summary and four future-interest statements to
  the final prediction. The action-adventure interest retrieves Uncharted 4
  at rank 1; the other three interests miss at cutoff 50. The direct SID beam
  ranks the target SID second.}
  \label{fig:case-study-both-hit}
\end{figure*}

\begin{figure}[t]
  \centering
  \includegraphics[width=0.9\columnwidth]{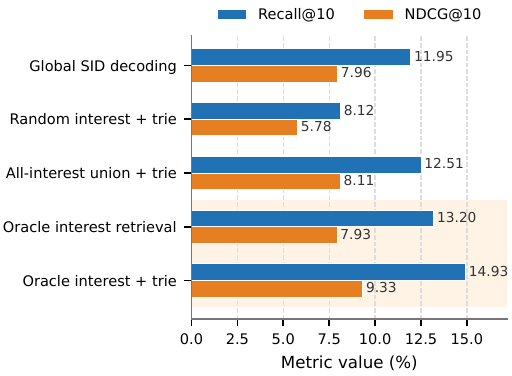}
  \caption{Interest-conditioned SID decoding on Video Games. Each interest
  retrieves its top-50 candidates. The all-interest variant merges and
  deduplicates candidates from all generated queries. The oracle uses the
  target rank to select the best generated query and therefore represents an
  upper bound.}
  \Description{Horizontal bars compare Recall at 10 and NDCG at 10 for global
  SID decoding, a random generated interest, the union of all generated
  interests, oracle-selected interest retrieval, and oracle-selected
  interest-conditioned SID decoding.}
  \label{fig:interest-conditioned-decoding}
\end{figure}

\paragraph{Potential of interest conditioned SID decoding.}
Figure~\ref{fig:interest-conditioned-decoding} compares direct SID decoding
with candidate-constrained variants on Video Games. A trie stores candidate
SIDs by prefix. At each decoding step, the model can choose only a token that
continues one of these candidate SIDs.

The random variant selects one generated interest and builds a trie from its
top-50 candidates. The all-interest variant merges and deduplicates the top-50
candidates from every generated interest. The oracle uses the held-out target
rank to choose the best interest already produced by the model. We evaluate
its cosine-ranked top 10 directly, then use its top-50 candidates to constrain
SID beam search.

A random interest performs substantially worse than global SID decoding. Merging all interests recovers this loss and slightly exceeds global SID decoding on both metrics, showing that the generated interests provide complementary candidate coverage despite differing in individual utility. Oracle retrieval improves recall, while SID decoding over the oracle trie gives the strongest recall and ranking. The random and oracle variants use the same pool size, so their gap comes from which interest is selected. This upper bound shows the potential of interest conditioned SID decoding. Selecting an interest without the target remains a promising direction for future work.

\subsection{Sensitivity to Reward Encoder and Cutoff}
\label{sec:reward-sensitivity}

Figure~\ref{fig:reward-sensitivity} varies the Qwen3-Embedding encoder size at
$K_{\mathrm{train}}=50$ and the training cutoff with the 4B encoder. All runs
share the same Stage-2 initialization and Stage-3 configuration, and all are
evaluated at I-R@50. The 4B encoder improves over 0.6B and matches 8B while
using half the online encoder capacity. This suggests that reward quality
saturates before the largest encoder.

Recommendation accuracy changes little across cutoffs. Interest Recall
increases with $K_{\mathrm{train}}$, while the SID metrics are highest at 50
in this sweep. The divergence shows that maximizing retrieval coverage does
not necessarily provide the best signal for SID learning. A larger cutoff
accepts more distant matches under the same binary reward, whereas a smaller
cutoff is more selective. We use $K_{\mathrm{train}}=50$ as the middle ground
and retain the 4B encoder.

\subsection{Qualitative Case Study}
\label{sec:case-study}

Figure~\ref{fig:case-study-both-hit} shows one Video Games test case in which
both paths find the target. Only the action-adventure interest retrieves
\textit{Uncharted 4: A Thief's End}, at rank 1. The other interests miss,
while the direct SID beam ranks the target second.

%% file: sections/conclusion.tex
\section{Conclusion}
\label{sec:conclusion}



We studied credit assignment in reasoning-enhanced SID recommendation, where a common issue is that the reward computed from the decoded item can be refined only within the SID span: it cannot separate rollouts that receive the same reward, and it carries no information when that reward is constant within a group. We proposed retrieval-grounded query attribution, which executes the generated interests against a frozen catalog retriever and routes retrieval advantage only to the interest spans that retrieve the target, leaving the final SID positions under item correctness. Across three Amazon Reviews datasets, the proposed method improves SID recommendation over a matched SID-and-trace control and restores group-relative learning signals in many SID-inactive groups. Ablations show that hit-aware routing outperforms coarser credit scopes, and the generated queries complement direct SID beam search. More broadly, these results show that intermediate predictions, once made executable, can supply localized supervision without a separate process reward model.

%% file: sections/ethical_considerations.tex
\section{Ethical Considerations}
\label{sec:ethical-considerations}

Our experiments use the public Amazon Reviews benchmark and do not collect new
user data. Nevertheless, methods that infer future interests from interaction
histories may expose sensitive preferences if deployed on identifiable logs.
Generated interests may also reproduce popularity and representation biases
in catalog metadata and the retriever, which can narrow user exposure and
reinforce filter bubbles. The generated trace is optimized for retrieval
utility, not verified as a faithful explanation of model behavior, and should
not be presented to users without further validation. A deployment should therefore restrict access to histories, avoid storing generated queries, and audit retrieval rewards across user and item groups.